\documentclass[preprint,showpacs,preprintnumbers,amsmath,amssymb]{revtex4}

\usepackage{graphicx}
\usepackage{dcolumn}
\usepackage{bm}

\begin{document}

\preprint{APS/123-QED}

\title{Similarity between Ni- and Zn-impurity effects on the superconductivity and Cu-spin correlation in La-214  high-$T_{\rm c}$ cuprates: Review based on the hole trapping by Ni}

\author{Y. Tanabe}\thanks{\\ Present address: WPI-Advanced Institute of Materials Research, Tohoku University, 6-3 Aoba, Aramaki, Aoba-ku, Sendai 980-8579, Japan}

\author{T. Adachi}\thanks{Corresponding author: adachi@teion.apph.tohoku.ac.jp}

\author{K. Suzuki}

\author{T. Kawamata$^1$}\thanks{\\ Present address:Department of Applied Physics, Graduate School of Engineering, Tohoku University, 6-6-05 Aoba, Aramaki, Aoba-ku, Sendai 980-8579, Japan}

\author{Risdiana$^{1, 2}$}

\author{T. Suzuki$^1$}\thanks{\\ Present address: Graduate School of Arts and Science, International Christian University, 3-10-2, Osawa, Mitaka, Tokyo 181-8585, Japan}

\author{I. Watanabe$^1$}

\author{Y. Koike}

\affiliation{Department of Applied Physics, Graduate School of Engineering, Tohoku University, 6-6-05 Aoba, Aramaki, Aoba-ku, Sendai 980-8579, Japan }

\affiliation{$^1$Advanced Meson Science Laboratory, Nishina Center for Accelerator-Based Science, RIKEN, 2-1 Hirosawa, Wako 351-0198}

\affiliation{$^2$Department of Physics, Faculty of Mathematics and Natural Sciences, Padjadjaran University, Jl. Raya Bandung-Sumedang Km.21 Jatinangor, Sumedang 45363, Indonesia}

\date{\today}

\begin{abstract}
Ni-substitution effects on the superconductivity and Cu-spin correlation have been investigated in La$_{2-x}$Sr$_x$Cu$_{1-y}$Ni$_y$O$_4$ from the electrical resistivity and muon spin relaxation measurements, taking into account the hole trapping by Ni recently suggested.
It has been found that the Ni-substitution suppresses the superconductivity, induces the localization of holes and develops the Cu-spin correlation to the $same$ $degree$ as the Zn substitution.
These suggest that Ni with a trapped hole tends to give rise to potential scattering of holes in the CuO$_2$ plane to the same degree as Zn, being discussed in relation to the so-called dynamical stripe correlations of spins and holes.

\end{abstract}

\pacs{74.62.Dh, 74.72.Gh, 76.75.+i, 74.25.F-}
\maketitle

\section{Introduction}
Study of impurity effects has been one of major ways for elucidating the mechanism of superconductivity.
In conventional $s$-wave superconductors, the suppression of the superconductivity by magnetic impurities is more marked than by non magnetic impurities. \cite{Tinkham}
In the hole-doped high-$T_{\rm c}$ cuprates, it is well known that the superconductivity is suppressed by the non magnetic impurity Zn more markedly than by the magnetic impurity Ni. \cite{Xiao}
Both nuclear magnetic/quadrupole resonance (NMR/NQR) experiments in La$_{2-x}$Sr$_x$Cu$_{1-y}$Zn$_y$O$_4$ (LSCZO) and La$_{2-x}$Sr$_x$Cu$_{1-y}$Ni$_y$O$_4$ (LSCNO) \cite{Ishida2, Kitaoka} and scanning tunneling microscope (STM) experiments in Bi$_2$Sr$_2$CaCu$_{2-y}$(Zn,Ni)$_y$O$_{8+\delta}$ \cite{Pan, Hudson} have revealed that the phase shift due to potential scattering by Zn is larger than by Ni.
As to the Cu-spin state, it has been found from muon spin relaxation ($\mu$SR) experiments in LSCZO and LSCNO that the Cu-spin correlation tends to be developed by Zn more effectively than by Ni. \cite{Watanabe1, Adachi1, Adachi2, Adachi3}
These results suggest that Ni-substitution effects on the electronic and Cu-spin states are weaker than Zn-substitution effects.

Recently, it has been suggested from neutron scattering, \cite{Matsuda, Hiraka1} magnetic susceptibility, \cite{Machi} x-ray absorption fine structure (XAFS), \cite{Hiraka2} $\mu$SR \cite{Tanabe2} measurements and a theoretical work using the exact diagonalization calculation \cite{Tsutsui} that a hole is bound around a Ni$^{\rm 2+}$ ion in the CuO$_2$ plane, leading to the decrease of the effective hole concentration.
Moreover, our specific heat, electrical resistivity, and magnetic susceptibility in La$_{2-x}$Sr$_x$Cu$_{1-y}$Ni$_y$O$_4$ have revealed that the strongly bound state of holes around Ni$^{2+}$ accompanied by the magnetically ordered state emerges at low temperatures in the underdoped regime and that it gradually changes to the Kondo state between Ni$^{2+}$ spins and holes above the effective hole concentration $p_{\rm eff}$, defined as $x$ - $y$, taking into account the hole trapping by Ni, = 0.13. \cite{Tanabe, szkn}
Therefore, Ni-substitution effects on the electronic and Cu-spin states in the high-$T_{\rm c}$ cuprates have to be reconsidered taking into account the hole trapping by Ni.

In this paper, in order to investigate Ni-substitution effects on the electronic and Cu-spin states on the basis of the hole trapping by Ni, we have performed electrical resistivity and $\mu$SR measurements in LSCNO. 
Moreover, we have compared the present results with previous results on Zn-substitution effects in LSCZO.

\section{Experimental}
Polycrystalline samples of LSCNO with $x$ = 0.08 - 0.18 and $y$ = 0 - 0.05 were prepared by the ordinary solid-state reaction method.
The details are described in our previous papers. \cite{Adachi2, Adachi3}
All of the samples were checked by the powder x-ray diffraction measurements to be of the single phase.
The electrical-resistivity measurements were carried out using the four-probe method to investigate the electronic state.
Zero-field (ZF) $\mu$SR measurements were performed at the Paul Scherrer Institute (PSI) in Switzerland using a continuous muon beam and at the RIKEN-RAL Muon Facility at the Rutherford-Appleton Laboratory in the UK using a pulsed muon beam.
The asymmetry parameter $A$($t$) at a time $t$ was given by $A$($t$) = $\{$$F$($t$) - $\alpha$$B$($t$)$\}$/$\{$$F$($t$) + $\alpha$$B$($t$)$\}$, where $F$($t$) and $B$($t$) are total muon events of the forward and backward counters, respectively.
The $\alpha$ is a calibration factor reflecting relative counting efficiencies between the forward and backward counters.
The $\mu$SR time spectrum, namely, the time evolution of $A$($t$) was measured down to 2 K to monitor the evolution of the Cu-spin correlation.

\section{Results}

Typical results of the temperature dependence of the electrical resistivity, $\rho$, divided by $\rho$ at 250 K, $\rho$$_{\rm N}$($T$), for LSCNO with $p_{\rm eff}$ = 0.08 and $y$ = 0 - 0.03, $p_{\rm eff}$ = 0.10 and $y$ = 0 - 0.03, and $p_{\rm eff}$ = 0.13 and $y$ = 0 - 0.05 are shown in Fig. 1, as well as those for LSCZO with $x$ = 0.10 and $y$ = 0 - 0.03.
For Ni-free samples with $y$ = 0, $\rho$$_{\rm N}$($T$) exhibits a local minimum around 50 - 60 K due to the localization of holes at low temperatures and then drops due to the superconducting (SC) transition with $p_{\rm eff}$ = 0.08 and 0.10, while a metallic $\rho$$_{\rm N}$($T$) is observed down to the SC transition temperature $T_{\rm c}$ with $p_{\rm eff}$ = 0.13.
Here we define $T_{\rm loc}$ as the temperature where the local minimum is observed and $T_{\rm c}$ at the midpoint of the SC transition in the $\rho$$_{\rm N}$($T$) vs. $T$ plot.
For Ni-substituted samples, $T_{\rm c}$ decreases with increasing $y$ and disappears around $y$ = 0.01 - 0.02 for $p_{\rm eff}$ = 0.08 and 0.10 and at $y$ = 0.05 for $p_{\rm eff}$ = 0.13.
The $T_{\rm loc}$ tends to increase with increasing $y$ for $p_{\rm eff}$ = 0.08 and 0.10, while $T_{\rm loc}$ is observed only in $y$ = 0.05 for $p_{\rm eff}$ = 0.13.
As for the Zn-substituted samples with $x$ = 0.10, $T_{\rm c}$ decreases with increasing $y$ and disappears for $y$ $>$ 0.01, which is almost the same as in the Ni-substituted samples with $p_{\rm eff}$ = 0.10.
Moreover, $T_{\rm loc}$ tends to increase with increasing $y$ in the Zn-substituted samples with $x$ = 0.10, which is also similar to that in the Ni-substituted samples with $p_{\rm eff}$ = 0.10.

Figures 2(a)-(c) show the ZF-$\mu$SR time spectra for LSCNO with $p_{\rm eff}$ = 0.08 and $y$ = 0 - 0.02, $p_{\rm eff}$ = 0.10 and $y$ = 0 - 0.03, and $p_{\rm eff}$ = 0.13 and $y$ = 0 - 0.05 at 2 K, respectively.
For comparison, the ZF-$\mu$SR time spectra for LSCZO with $x$ = 0.10 and $y$ = 0 - 0.03 at 2 K are also shown in Fig. 2(d). \cite{Adachi1}
The inset in Fig. 2(b) shows the ZF-$\mu$SR time spectrum for the Ni-free sample with $p_{\rm eff}$ = 0.10 at 20 K.
This shows Gaussian-like slow depolarization, originated from the nuclear dipole field randomly distributed at the muon site.
This means no effects of Cu spins coming into the $\mu$SR time (frequency) window.
Almost the same spectrum has been obtained at 20 K in every sample.
At 2 K, for Ni-free samples with $p_{\rm eff}$ = 0.08 and 0.10, fast depolarization of muon spins is observed, indicating the slowing down of the Cu-spin fluctuations, while almost no fast depolarization of muon spins is observed for the Ni-free sample with $p_{\rm eff}$ = 0.13.
For Ni-substituted samples, a muon-spin precession due to the emergence of a magnetically ordered state of Cu spins is observed with $p_{\rm eff}$ = 0.08 and 0.10.
For $p_{\rm eff}$ = 0.13, only fast depolarization of muon spins is observed at $y$ = 0.02 - 0.05 and no precession is observed.
As for the Zn-substituted samples with $x$ = 0.10, a muon spin precession is observed, which is similar to the Ni-substituted samples with $p_{\rm eff}$ = 0.10.

In order to investigate details of the Cu-spin state, the $\mu$SR time spectra were analyzed using the following three-component function:
\begin{equation}
A(t) = A_0 e^{-\lambda_0t}G_Z(\Delta,t) + A_1 e^{-\lambda_1t} + A_2 e^{-\lambda_2t}{\rm cos}(\omega t + \phi).
\label{eq1}
\end{equation}
The first term represents the slowly depolarizing component in a region where Cu spins fluctuate fast beyond the $\mu$SR frequency range ($10^6 - 10^{11}$ Hz). 
The $A_0$ and $\lambda_0$ are the initial asymmetry and depolarization rate of the slowly depolarizing component, respectively. 
The $G_Z(\Delta,t)$ is the static Kubo-Toyabe function dependent on $\Delta$ describing the half width of the distribution of the nuclear dipole field at the muon site. \cite{Uemura}
The second term represents the fast depolarizing component in a region where Cu-spin fluctuations slow down and/or a short-range magnetic order is formed. 
The $A_1$ and $\lambda_1$ are the initial asymmetry and depolarization rate of the fast depolarizing component, respectively.
The third term represents the muon-spin precession in a region where a long-range magnetic order is formed. 
The $A_2$ is the initial asymmetry. 
The $\lambda_2$, $\omega$ and $\phi$ are the damping rate, frequency and phase of the muon-spin precession, respectively. 
Here, $A$($t$) is the normalized asymmetry and  $A_0$, $A_1$ and $A_2$ are normalized values, that is, $A_0$ + $A_1$ + $A_2$ = 1.
The time spectra are well fitted with Eq. (1), as shown by the solid lines in Fig. 2.

The temperature dependence of $A_0$ is often used to monitor the magnetic transition, because it reflects the volume fraction of a non magnetic region. \cite{Torikai, Watanabe4, Watanabe5}
At $A_0$ = 1, the $\mu$SR time spectrum is represented only with the first term of Eq. (1), indicating that all the Cu spins fluctuate fast beyond the $\mu$SR frequency window.
On the contrary, the saturation of $A_0$ around 1/3 means that all the Cu spins are in a short- or long-range magnetically ordered state.
Typical results of the temperature dependence of $A_0$ for LSCNO with $p_{\rm eff}$ = 0.08 are shown in Fig. 3.
For the Ni-free sample with $y$ = 0, $A_0$ decreases with decreasing temperature below 4 K in correspondence to the emergence of a fast depolarization of muon spins shown in Fig. 2, indicating that the decrease of $A_0$ monitors the development of the Cu-spin correlation.
For the Ni-substituted sample with $y$ = 0.01, both the saturation of $A_0$ and the muon-spin precession shown in Fig. 2 are observed at 2 K, indicating that the saturation of $A_0$ at low temperatures monitors the emergence of a magnetically ordered state.
Accordingly, we define the magnetic transition temperature, $T_{\rm N}$, at the midpoint of the change in the normalized $A_0$ from unity to the saturated value of $A_0$ at 2 K in the magnetically ordered state. \cite{Adachi1}

Figure 4 shows dependencies of $T_{\rm c}$, $T_{\rm loc}$, $T_{\rm N}$ and the internal magnetic field at the muon site, $H_{\rm int}$, at 2 K on the Ni concentration $y$ for LSCNO with $p_{\rm eff}$ = 0.08, 0.10, and 0.13.
The $H_{\rm int}$ is defined as $H_{\rm int}$ = $\omega$/$\gamma$$_{\rm \mu}$ where $\gamma$$_{\rm \mu}$ is the gyromagnetic ratio of muon spin ($\gamma$/2$\pi$ = 13.55 MHz/kOe).
For comparison, data for LSCZO with $x$ = 0.08, 0.10 and 0.13 are also shown. \cite{Adachi1}
As for the superconductivity, it is found that $T_{\rm c}$ decreases with increasing $y$ and disappears around $y$ = 0.01 - 0.02 for LSCNO with $p_{\rm eff}$ = 0.08 and 0.10.
To our surprise, the dependences of $T_{\rm c}$ on $y$ for LSCNO with $p_{\rm eff}$ = 0.08 and 0.10 are almost identical to those for LSCZO with $x$ = 0.08 and 0.10, respectively.
For LSCNO with $p_{\rm eff}$ = 0.13, $T_{\rm c}$ disappears at $y$ = 0.05, while it does at $y$ = 0.02 for LSCZO with $x$ = 0.13.

As for the electronic state in the normal state, it is found that $T_{\rm loc}$ increases with increasing $y$ for LSCNO with $p_{\rm eff}$ = 0.08 and 0.10.
Dependences of $T_{\rm loc}$ on $y$ for LSCNO with $p_{\rm eff}$ = 0.08 and 0.10 are in rough agreement with those for LSCZO with $x$ = 0.08 and 0.10, respectively.
For LSCNO with $p_{\rm eff}$ = 0.13, $T_{\rm loc}$ is observed only for $y$ = 0.05, while $T_{\rm loc}$ tends to increase with increasing $y$ for LSCZO with $x$ = 0.13.

As for the Cu-spin correlation, it is found that both $T_{\rm N}$ and $H_{\rm int}$ are developed through only 0.5 $\%$ Ni substitution for LSCNO with $p_{\rm eff}$ = 0.08.
For LSCNO with $p_{\rm eff}$ = 0.10, both $T_{\rm N}$ and $H_{\rm int}$ tend to show the maximum at $y$ = 0.01 - 0.03.
To our surprise, values of $H_{\rm int}$ for LSCNO with $p_{\rm eff}$ = 0.10 are almost identical to those for LSCZO with $x$ = 0.10.
Moreover, the $y$ dependence of $T_{\rm N}$ for LSCNO with $p_{\rm eff}$ = 0.10 is similar to that for LSCZO with $x$ = 0.10.
For LSCNO with $p_{\rm eff}$ = 0.13, neither $T_{\rm N}$ nor $H_{\rm int}$ are observed above 2 K, while both $T_{\rm N}$ and $H_{\rm int}$ show the maximum at $y$ = 0.0075 - 0.02 and then decrease with increasing $y$ and disappear above $y$ = 0.05 for LSCZO with $x$ = 0.13.

\section{Discussion}


\subsection{Similarity between Ni- and Zn-substitution effects in $p_{\rm eff}$ = 0.08 and 0.10} 
It has been found from $\rho$ measurements for LSCNO with $p_{\rm eff}$ = 0.08 and 0.10 that $T_{\rm c}$ decreases and $T_{\rm loc}$ increases with increasing $y$.
Compared with Zn-substituted samples, dependences of $T_{\rm c}$ on $y$ for LSCNO with $p_{\rm eff}$ = 0.08 and 0.10 are almost identical to those for LSCZO with $x$ = 0.08 and 0.10, respectively.
Moreover, dependences of $T_{\rm loc}$ on $y$ for LSCNO with $p_{\rm eff}$ = 0.08 and 0.10 are in rough agreement with those for LSCZO with $x$ = 0.08 and 0.10, respectively.
These results suggest that the Ni substitution suppresses the superconductivity and induces the localization of holes in $p_{\rm eff}$ = 0.08 and 0.10 to the same degree as the Zn substitution in $x$ = 0.08 and 0.10.
These are in sharp contrast to reported impurity effects in conventional BCS superconductors and high-$T_{\rm c}$ cuprates. \cite{Tinkham, Xiao, Ishida2, Kitaoka, Hudson}
Here, it is noted that although the intergrain coupling in polycrystalline samples may affect the value of $\rho$ to some extent, the systematic changes in $T_{\rm loc}$ on $y$ suggest that the present results reflect the intrinsic nature of the electronic transport in the CuO$_2$ plane.
As for the Cu-spin correlation, the $\mu$SR measurements for LSCNO with $p_{\rm eff}$ = 0.08 have revealed that both $T_{\rm N}$ and $H_{\rm int}$ are developed through only 0.5 $\%$ Ni substitution.
For $p_{\rm eff}$ = 0.10, both $T_{\rm N}$ and $H_{\rm int}$ are also developed through only 0.5 $\%$ Ni substitution and show the maximum at $y$ = 0.01 - 0.03.
Compared with Zn-substituted samples, values of $H_{\rm int}$ for LSCNO with $p_{\rm eff}$ = 0.10 are almost identical to those for LSCZO with $x$ = 0.10.
Moreover, the $y$ dependence of $T_{\rm N}$ for LSCNO with $p_{\rm eff}$ = 0.10 is similar to that for LSCZO with $x$ = 0.10.
These results suggest that a small amount of Ni develops the Cu-spin correlation in $p_{\rm eff}$ = 0.10 to the same degree as a small amount of Zn in $x$ = 0.10
These are also in sharp contrast to previous reports of impurity effects on the Cu-spin correlation in high-$T_{\rm c}$ cuprates. \cite{Adachi2, Adachi3}.
Accordingly, taking into consideration the hole-trapping effect by Ni, it has been clarified that the Ni substitution suppresses the superconductivity and develops the localization of holes and the Cu-spin correlation to the same degree as the Zn substitution.

Here we discuss the quantitative similarity between impurity effects of magnetic Ni and non-magnetic Zn.
Supposing the hole trapping by Ni, the effective valence of Ni with a trapped hole is close to being trivalent, leading to the enhancement of the potential scattering of holes by Ni with a trapped hole.
Moreover, the effective value of the spin quantum number $S$ of Ni with a trapped hole is regarded as 1/2, which is the same as that of Cu$^{2+}$ spins, indicating the magnetic scattering by Ni being weakened.
In $d$-wave superconductors, it is known that the phase shift due to the potential scattering suppresses the superconductivity. \cite{Hirschfeld, Miyake, Hotta}
In the underdoped high-$T_{\rm c}$ cuprates, the localization of holes may break the coherency between Cooper pairs owing to the small SC carrier density.
The localization of holes also develops the Cu-spin correlation.
Therefore, one possible reason for the quantitative similarity between Ni and Zn impurity effects is that the strong potential scattering by Ni owing to the pseudo trivalent state of Ni suppresses the superconductivity and develops the localization of holes and Cu-spin correlation to the same degree as that by Zn which has a closed shell of 3$d$ orbitals.

Another possible reason for the quantitative similarity between Ni and Zn impurity effects is due to the development of stripe correlations of spins and holes around Ni and Zn.
Former $\mu$SR experiments in LSCZO have revealed that a small amount of Zn tends to stabilize the Cu-spin correlation around Zn, suggesting the development of stripe correlations of spins and holes around Zn. \cite{Watanabe1, Adachi1, Adachi3}
Recent neutron scattering measurements have revealed that a static charge stripe order is induced through only 1 $\%$ Zn substitution in LSCZO with $x$ = 0.12. \cite{Fujita}
The $\rho$ measurements for LSCZO have also suggested that the upturn of $\rho$ at low temperatures is not due to simple localization of holes but due to the emergence of a static charge stripe order. \cite{Komiya}
The present results have revealed that a small amount of Ni develops the localization of holes and the Cu-spin correlation to the same degree as a small amount of Zn.
Therefore, it is likely that a small amount of Ni develops the stripe correlations of spins and holes to the same degree as a small amount of Zn, leading to the suppression of the superconductivity, the localization of holes, and the development of the Cu-spin correlations.

What is the origin of the stabilization of the dynamical stripe correlations induced by Ni?
It has been found from neutron scattering measurements in La$_{2-x}$Sr$_x$Cu$_{1-y}$M$_y$O$_4$ (M = Ga, Zn) that Ga$^{\rm 3+}$ tends to stabilize the dynamical stripe correlations to the same degree as Zn$^{\rm 2+}$. \cite{Fujita}
Based on the hole trapping by Ni, the effective valence of Ni with a trapped hole is close to being trivalent, which is the same as that of Ga$^{\rm 3+}$.
Therefore, it is possible that the quasi trivalent state of Ni owing to the hole trapping stabilizes the dynamical stripe correlations to the same degree as Zn.

\subsection{Discrepancy between Ni- and Zn-substitution effects with $p_{\rm eff}$ = 0.13}
It has been found from $\rho$ measurements for LSCNO with $p_{\rm eff}$ = 0.13 that $T_{\rm c}$ decreases with increasing $y$ and disappears at $y$ = 0.05 and that $T_{\rm loc}$ is observed only in $y$ = 0.05.
Compared with Zn-substituted samples, both the decrease of $T_{\rm c}$ and the increase of $T_{\rm loc}$ induced by Ni in $p_{\rm eff}$ = 0.13 are weaker than those induced by Zn.
These results suggest that both the suppression of the superconductivity and the localization of holes induced by Ni in $p_{\rm eff}$ = 0.13 are weaker than those by Zn in $x$ = 0.13.
As for the Cu-spin correlation, the $\mu$SR measurements have revealed that fast depolarization of muon spins is observed at 2 K for Ni-substituted samples with $p_{\rm eff}$ = 0.13, \cite{Tanabe2} while no fast depolarization of muon spins is observed at 2 K for Ni-free samples with $x$ $\geq$ 0.13, \cite{Adachi1, Adachi3, Risdiana} suggesting that the Ni substitution develops the Cu-spin correlation in $p_{\rm eff}$ = 0.13.
Compared with Zn-substituted samples, neither $T_{\rm N}$ nor $H_{\rm int}$ are observed above 2 K for Ni-substituted samples with $p_{\rm eff}$ = 0.13, while both $T_{\rm N}$ and $H_{\rm int}$ are developed for Zn-substituted samples with $x$ = 0.13, suggesting that the development of the Cu-spin correlation induced by Ni in $p_{\rm eff}$ = 0.13 is weaker than that induced by Zn in $x$ = 0.13.
Taking into consideration the stripe correlations of spins and holes, it appears that the Ni substitution develops the dynamical stripe correlations in $p_{\rm eff}$ = 0.13, but the development of the dynamical stripe correlations induced by Ni in $p_{\rm eff}$ = 0.13 is weaker than that induced by Zn in $x$ = 0.13.
It has been found from specific-heat and $\rho$ measurements for LSCNO with $p_{\rm eff}$ $\geq$ 0.13 that the electronic specific heat divided by temperature is enhanced on account of the Kondo effect due to Ni spins. \cite{Tanabe, szkn}
In this case, the excess magnetic moment of Ni$^{2+}$ spins ($S$ = 1) is perfectly screened only at the ground state, meaning that the actual hole concentration in $p_{\rm eff}$ = 0.13 tends to be larger than $p_{\rm eff}$ except for the ground state.
It has been suggested from $\mu$SR measurements for LSCZO that the impurity-induced stabilization of the dynamical stripe correlations tends to be weakened gradually with increasing hole concentration in the overdoped regime. \cite{Risdiana}
Accordingly, it appears that the stabilization of the dynamical stripe correlations is weakened in $p_{\rm eff}$ = 0.13 due to the weakening of the binding of holes around Ni, leading to the weakening of both the suppression of the superconductivity, the localization of holes and the stabilization of the Cu-spin correlation induced by Ni with $p_{\rm eff}$ = 0.13.

\section{Summary}
Based on the hole trapping by Ni, we have investigated Ni-substitution effects on the superconductivity and Cu-spin correlation from $\rho$ and $\mu$SR measurements in La$_{2-x}$Sr$_x$Cu$_{1-y}$Ni$_y$O$_4$.
For $p_{\rm eff}$ = 0.08 and 0.10, it has been found that the Ni substitution suppresses the superconductivity and induces the localization of holes to the same degree as the Zn substitution for $x$ = 0.08 and 0.10, respectively.
Moreover, the Ni substitution develops the Cu-spin correlation for $p_{\rm eff}$ = 0.10 to the same degree as the Zn substitution for $x$ = 0.10.
These suggest that Ni with a trapped hole tends to give rise to potential scattering of holes in the CuO$_2$ plane to the same degree as Zn, which may be related to the development of dynamical stripe correlations of spins and holes.
By contrast, for $p_{\rm eff}$ = 0.13 where a Kondo effect due to Ni spins is observed in the specific-heat measurements, both the suppression of the superconductivity, the localization of holes and the development of the Cu-spin correlation induced by Ni are weaker than those by Zn in $x$ = 0.13.
This may be due to the actual hole concentration being larger than $p_{\rm eff}$ in the Ni-substituted sample owing to the weak binding of holes by Ni in the Kondo state.

\section{Acknowledgements}
We thank A. Amato and R. Sheuermann at PSI for their technical support in the $\mu$SR measurements. 
The $\mu$SR measurements at PSI were partially supported by KEK-MSL Inter-University Program for Oversea Muon Facilities, by the Global COE Program gMaterials Integration (International Center of Education and Research), Tohoku University,h MEXT, Japan, and by gEducation Program for Biomedical and Nano-Electronics, Tohoku Universityh Program for Enhancing Systematic Education in Graduate Schools, MEXT, Japan.
Y. T. was supported by the Japan Society for the Promotion of Science.


\clearpage

\begin{figure}
\includegraphics[width=0.9\linewidth]{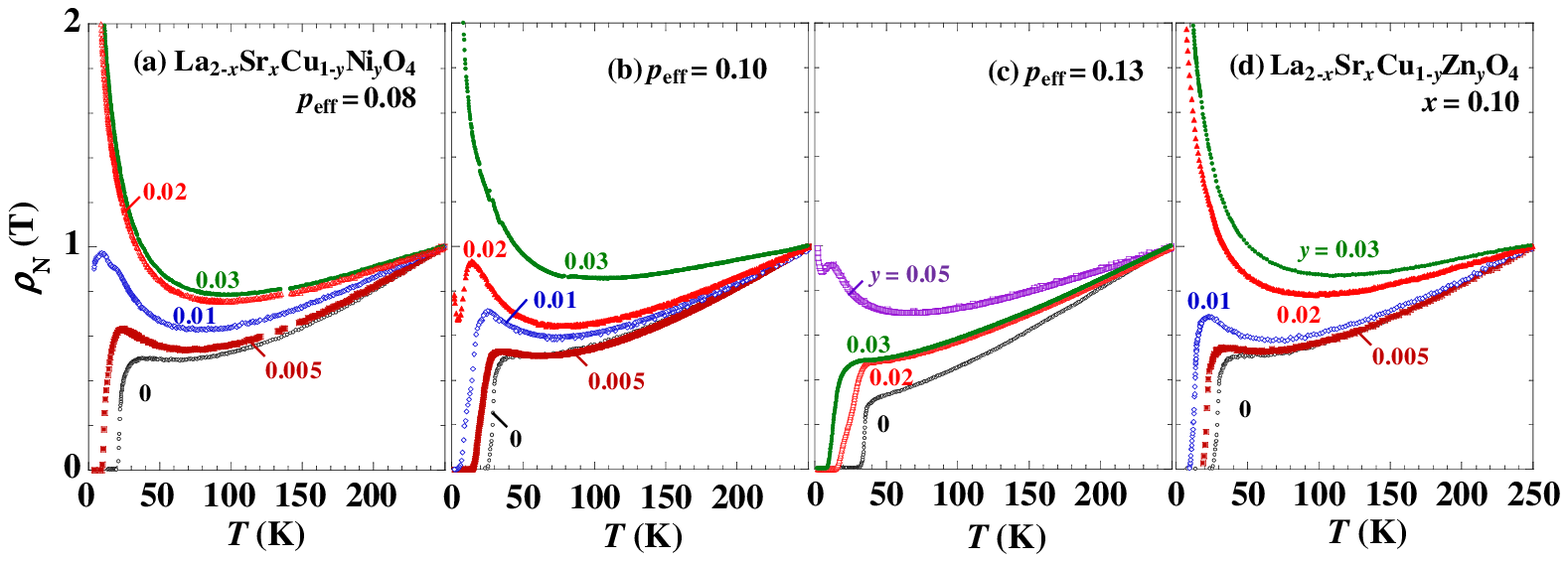}
\caption{(Color online) Temperature dependence of the electrical resistivity $\rho$ divided by $\rho$ at 250 K, $\rho$$_{\rm N}$($T$), for typical values of $y$ in La$_{2-x}$Sr$_x$Cu$_{1-y}$Ni$_y$O$_4$ with (a) $p_{\rm eff}$ = 0.08, (b) $p_{\rm eff}$ = 0.10 and (c) $p_{\rm eff}$ = 0.13.
For comparison, the temperature dependence of $\rho$$_{\rm N}$($T$) for La$_{2-x}$Sr$_x$Cu$_{1-y}$Zn$_y$O$_4$ with $x$ = 0.10 is shown in (d) (Ref. \cite{Adachi1}).}
\end{figure}

\clearpage

\begin{figure}
\includegraphics[width=0.4\linewidth]{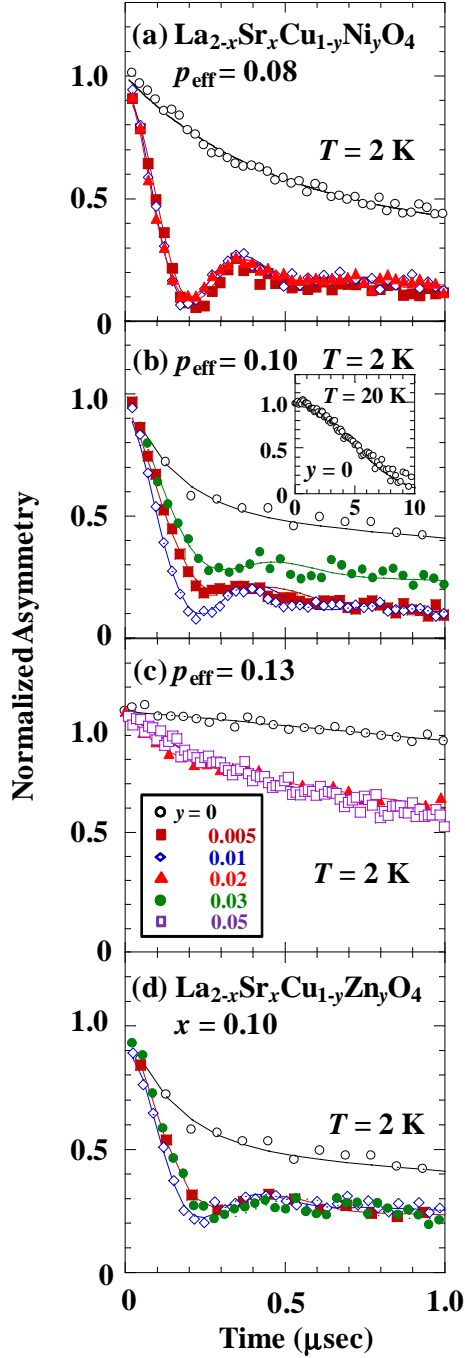}
\caption{(Color online) Zero-field $\mu$SR time spectra in an early time region from 0 to 1 $\mu$s at 2 K for La$_{2-x}$Sr$_x$Cu$_{1-y}$Ni$_y$O$_4$ with (a) $p_{\rm eff}$ = 0.08, (b) $p_{\rm eff}$ = 0.10, and (c) $p_{\rm eff}$ = 0.13.
The inset in (b) shows the zero-field $\mu$SR time spectrum at 20 K for La$_{2-x}$Sr$_x$Cu$_{1-y}$Ni$_y$O$_4$ with $p_{\rm eff}$ = 0.10 and $y$ = 0.
For comparison, zero-field $\mu$SR time spectra for La$_{2-x}$Sr$_x$Cu$_{1-y}$Zn$_y$O$_4$ with $x$ = 0.10 are also shown in (d) (Ref. \cite{Adachi1}).
Solid lines indicate the best-fit results using the analysis function : $A(t) = A_0 e^{-\lambda_0t}G_Z(\Delta,t) + A_1 e^{-\lambda_1t} + A_2 e^{-\lambda_2t}{\rm cos}(\omega t + \phi)$. }
\end{figure}

\clearpage

\begin{figure}
\includegraphics[width=0.5\linewidth]{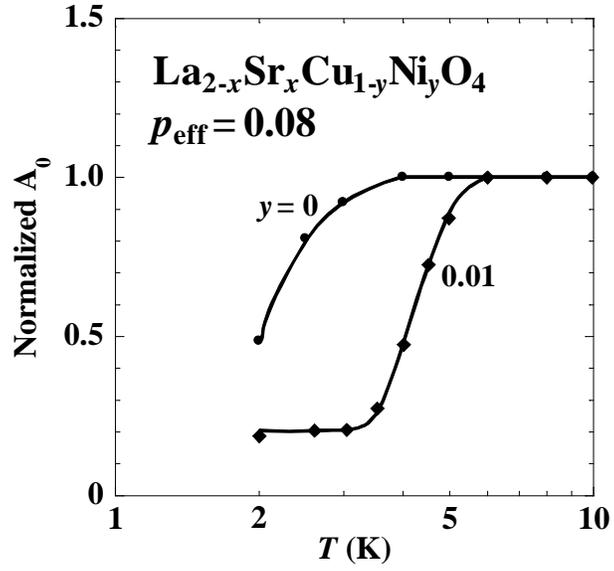}
\caption{Temperature dependence of the initial asymmetry of the slowly depolarizing component $A_0$ normalized by its value at 20 K in La$_{2-x}$Sr$_x$Cu$_{1-y}$Ni$_y$O$_4$ with $p_{\rm eff}$ = 0.08 and $y$ = 0, 0.01.
Solid lines are to guide the reader's eye.}
\end{figure}

\clearpage

\begin{figure}
\includegraphics[width=0.7\linewidth]{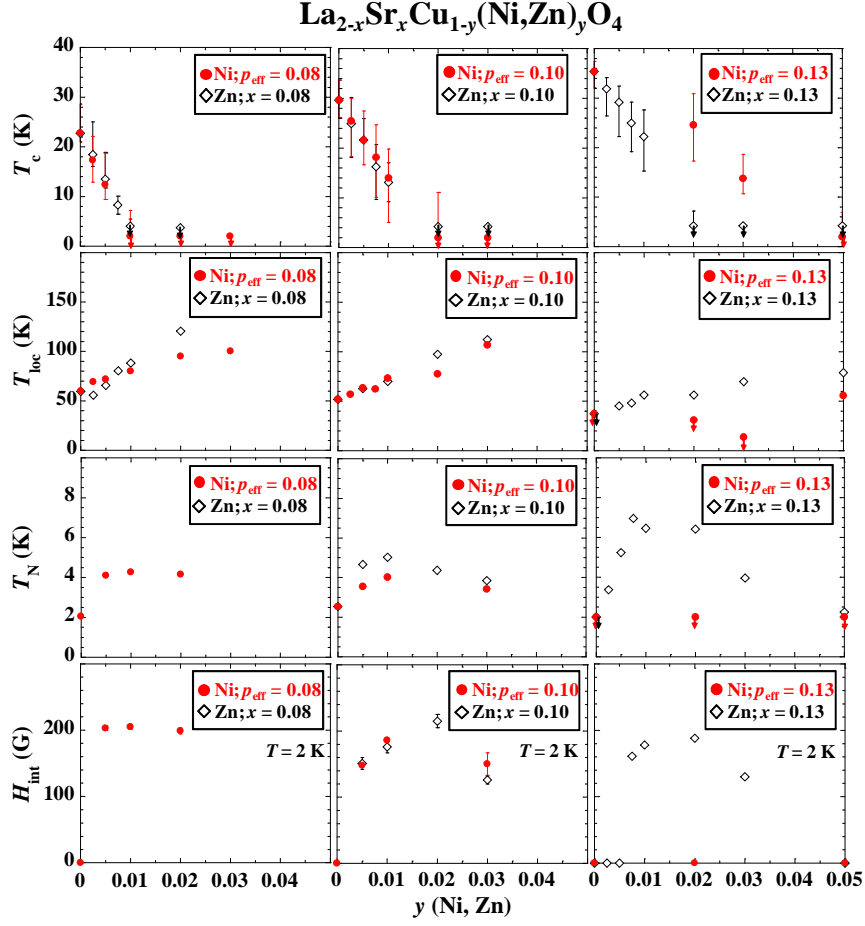}
\caption{(Color online) Dependences of the superconducting transition temperature $T_{\rm c}$, the hole-localization temperature $T_{\rm loc}$, the magnetic transtion temperature $T_{\rm N}$, and the internal magnetic field $H_{\rm int}$ at 2 K on the Ni concentration $y$ for La$_{2-x}$Sr$_x$Cu$_{1-y}$Ni$_y$O$_4$ with $p_{\rm eff}$ = 0.08, 0.10, and 0.13 are shown by red circles.
For comparison, data for La$_{2-x}$Sr$_x$Cu$_{1-y}$Zn$_y$O$_4$ with $x$ = 0.08, 0.10 and 0.13 are also shown by open diamonds (Ref. \cite{Adachi1}).
}
\end{figure}

\end{document}